\documentclass[10pt,a4paper]{article}

\usepackage{graphics}
\newcommand{\bea}{\begin{eqnarray}}
\newcommand{\eea}{\end{eqnarray}}
\newcommand{\be}{\begin{equation}}
\newcommand{\ee}{\end{equation}}

\def\be{\begin{eqnarray}}
\def\ee{\end{eqnarray}}
\def\bd{\begin{displaymath}}
\def\ed{\end{displaymath}}

\def\ga{\gamma}

\def\ADNDT{{At. Data Nucl. Data Tables }}

\def\NP{Nucl. Phys. }
\def\PR{Phys. Rev. }
\def\PRL{Phys. Rev. Lett. }

\def\jpg{J. Phys. G: Nucl. Part. Phys. }
\def\EPJ{{Eur. Phys. J. }}
\def\etal{{\em et al.}}

\begin{document}
\title{Relativistic Mean Field in $A\approx$80 nuclei and low energy proton 
reactions}

\author{Chirashree Lahiri and G. Gangopadhyay\\
%
Department of Physics, University of Calcutta\\
92, Acharya Prafulla Chandra Road, Kolkata-700 009, India}
\date{}
\maketitle
\begin{abstract}
Relativistic Mean Field calculations have been performed for a number of nuclei 
in mass  $A\approx$80 region. Ground state binding energy, charge radius and charge
density values have been compared with experiment. Optical potential have been 
generated folding the nuclear density with the microscopic nuclear interaction
DDM3Y. S-factors for  low energy ($p,\gamma$) and ($p,n$) reactions have
been calculated and compared with experiment.

\end{abstract}

Relativistic Mean Field (RMF) approach has proved to be very successful in 
explaining different features 
of stable and exotic nuclei like ground state binding energy, deformation, 
radius, excited states, spin-orbit splitting, neutron halo, etc\cite{RMF1}. 
Particularly, the radius and the nuclear density are known to be well 
reproduced.
Its has led to its application to nuclear
reactions also.

Low energy reactions are very important from the astrophysical point
of view. In astrophysical environments, neutron and proton reactions 
are the keys to nucleosynthesis of heavy elements. Mass 80 region is an 
interesting one as ($p,\gamma$) and ($n,p$) reactions play important roles 
in determining the abundance of elements. In this mass region, there are
some proton-rich naturally occurring isotopes of elements known as $p$
 nuclei which can not be produced via $s$- or $r$- process. Mainly, 
proton capture reactions contribute to the formation of such nuclei. Recent works [See {\em e.g.} 
\cite{astro1}] have emphasized the importance of a number of charge exchange 
reactions in this mass region for production of very light $p$ nuclei. The 
relevant astrophysical rates can directly be derived from the ($p,n$) data, though the 
target is in the ground state and reaction has  negative a Q-value\cite{Rb}.
A number of recent experiments has focussed on reactions using protons having a
few MeV energy in mass 80 region.

Calculation of isotopic abundance requires a network calculation involving 
many thousands of reactions. Despite the importance of ($p,\ga$) or ($p,n$) 
reactions in explaining the abundance of $p$ nuclei, experimental data are
rather scarce due unavailability of the target nuclei on 
earth. Thus, one often has to depend on theory for these reactions. The 
calculations essentially utilize the Hauser-Feshbach formalism where,  
the optical model potential, a key ingredient, is often 
taken in a local or a global form. It is also possible to use a 
microscopic optical potential constructed utilizing nuclear densities.
If the target is stable, the density of the nucleus is available through 
electron scattering. However, in absence of stable target, theory remains our 
sole guide to describe the density. Thus, it is imperative to test the 
theoretical calculations, where experimental data are available, to 
verify its applicability. We aim to to check the success of microscopic 
optical potentials based on mean field densities in explaining the 
reaction cross sections. A good description depending essentially on
theory will allow one to extend the present method to 
the critical reactions which are beyond present day laboratory capabilities.

Calculations using microscopic potentials have been able to explain the 
observed elastic scattering cross sections even in nuclei far off the 
stability valley (See, {\em e.g.} Ref. \cite{CBe} and references therein). Low
energy projectiles probe only the outermost regions of the target nuclei. 
Hence, the nuclear skin plays a very important role in such reactions. The 
density information should be available from theoretical 
calculations. This method has been utilized to study low energy proton 
capture reactions in Ni and Cu nuclei\cite{NiCu} and  nuclei in $A=60-80$
region\cite{6080}.

For the present study we have selected a number of low energy 
proton reactions for their astrophysical relevance. 
The reactions $^{84,86,87}$Sr($p,\ga$) were investigated
through activation technique by Gy\"{u}rky \etal\ in Ref. \cite{Sr1}. 
It is important to note that $^{84}$Sr is 
another $p$ nucleus. In beam measurements were performed by
Galanopoulos \etal\cite{Sr2} to find out the cross sections for the 
reaction  $^{88}$Sr($p,\ga$). 
As for charge exchange reactions, three reactions, identified as
important by Rapp \etal\cite{astro1} and for which experimental cross sections are available, 
have been selected for study. 
The reaction  $^{75}$As($p,n$) was studied in \cite{As1,As2,As3} 
through in beam detection of neutrons.
Finally, the reactions $^{76}$Ge($p,n$)\cite{Ge}  and  $^{85}$Rb($p,n$)\cite{Rb}
were studied by activation technique. In the present work, we  
investigate the reactions mentioned above in a microscopic approach.


Theoretical density profiles were extracted in the RMF 
approach.
There are different variations of the Lagrangian density as
well as a number of different parametrizations. In the present work we have 
employed the FSU Gold\cite{prl} Lagrangian density.
It  contains, apart 
from the usual terms for a nucleon meson system, nonlinear terms involving 
self coupling of scalar-isoscalar meson,  and additional terms describing 
self-coupling of the vector-isoscalar meson and coupling between the 
vector-isoscalar meson and the vector-isovector meson. 

Pairing has been introduced under the BCS approximation using a zero range 
pairing force of strength 300 MeV-$fm$ for both proton and neutrons.
The RMF+BCS equations are solved under 
the usual assumptions of classical meson fields, time reversal symmetry, no-sea 
contribution, etc. 
Since we need the densities in co-ordinate space, the Dirac and the Klein 
Gordon equations have directly been solved in that space.
This approach has earlier been used 
\cite{CBe,CaNi,Zr} in neutron rich nuclei in different mass regions. 

The microscopic optical model potentials for the reactions are obtained using 
effective interactions 
derived from the nuclear matter calculation in the local density approximation,
{\em i.e.} by substituting the nuclear matter density
with the density distribution of the finite nucleus.  
In the present work, the microscopic nuclear potentials have been constructed 
by 
folding the  density dependent DDM3Y\cite{ddm3y1,ddm3y2}
effective interaction with the
densities from the RMF calculation.
This interaction, obtained from a finite range energy independent M3Y 
interaction by adding a zero range energy dependent pseudopotential and 
introducing a density dependent factor, has been employed successfully in 
nucleon nucleus as well as nucleus nucleus scattering, calculation of proton 
radioactivity, etc. The density dependence has been chosen in the form 
$C(1-\beta\rho^{2/3})$\cite{ddm3y2}, the constants being obtained from a nuclear
matter calculation\cite{ddm3y3}. The real and the imaginary parts of the 
potential are taken as 0.7 times and 0.1 times the DDM3Y potential, 
respectively. This normalization have also been used in our earlier work
on $(p,\gamma)$ reactions in lighter nuclei\cite{6080}.
We have checked that the above values adequately describe the cross section
measurements. Of course, these parameters can be tuned to fit the cross 
sections in individual reactions. 
For example, in $^{84}$Sr, if we choose the imaginary part of the potential as 
0.3 times of the DDM3Y potential, the result will differ by 10\% and fit the
experimental data better.  However, we believe a single parametrization for the entire mass region to be 
more useful. 

The Coulomb potentials are  similarly constructed by folding 
the Coulomb interaction with the microscopic proton densities. 
We have already used such potentials to calculate life times for proton, alpha 
and cluster radioactivity\cite{alp} as well as elastic proton 
scattering\cite{CBe} in different mass regions of the periodic table.

Reaction calculations have been performed with the computer code TALYS 
1.2\cite{talys} assuming spherical 
symmetry for the target nuclei. DDM3Y interaction is not a standard input 
of TALYS but can easily be incorporated.
Though the nuclear matter-nucleon potential does not include a spin-orbit 
term, the code provides a spin-orbit potential from the Scheerbaum 
prescription\cite{SO} coupled with the phenomenological complex potential 
depths. The default form for this potential
given in the code has been used without any modification. 

The TALYS code has a number of other useful features. We have
employed the full Hauser-Feshbach calculation with transmission coefficients 
averaged over total angular momentum values and with corrections due to width 
fluctuations. Up to thirty discrete levels of the nuclei involved have been 
included in the calculation. 

Our calculations, being more microscopic, are
more restricting. Yet, the rate depends on the models of the level density
and the E1 gamma strength function adopted in the calculation of cross sections.
Phenomenological models are usually fine tuned for nuclei near the
stability valley. Microscopic prescriptions, on the other hand, can
be extended to the drip lines,  and hence, have been assumed in all nuclei.
We have calculated our results with microscopic level densities
in Hartree-Fock (HF) and Hartree-Fock-Bogoliubov (HFB) methods, calculated
in TALYS by Goriley and Hilaire, respectively. We have also compared our results
using phenomenological level densities from constant temperature Fermi gas 
model, back shifted Fermi gas model and generalised superfluid model from 
TALYS. 
The cross sections are very much dependent on the level density chosen,
sometimes changing by a factor of 50\%. 
We find that in most of the cases, the HFB densities fit the experimental data 
better in our formalism.

For E1 gamma strength functions, results derived from HF+BCS and HFB 
calculations, available in the TALYS data base,  have been employed. 
In agreement with our observation in in \cite{6080}, here also the 
results for HFB calculations describe the S-factors reasonably well
and we present our results for that approach only.

It is possible to scale the theoretical capture
cross sections to match with experiment
using a parameter $G_{norm}$ in the code used to scale the gamma-ray 
transmission coefficient. However, for the present paper,
we have not scaled the theoretical results.
All the parameters in the Lagrangian density and the interaction are 
standard ones and have not been changed.


As the density profile is the important 
quantity in our formalism, a comparison of the radii values can provide some 
idea about the agreement of the calculated densities with experiments. 
In Table 1, we compare our results for the binding energy and charge radii 
($r_{ch}$) with measurements for those nuclei in this mass region, which 
have been used as targets for low energy proton capture or charge exchange 
reactions.  The binding energy values from the mean
field approach have been corrected using the formalism developed in
\cite{rmfcor,rmfcor1}. The experimental binding energy values are from
Ref. \cite{audi}.

 Charge radii have been calculated from the charge densities, which, in  
turn, have been obtained from the calculated point proton
density $\rho_p$ by taking into account the finite size of the proton.
The point proton density is convoluted with a Gaussian form factor
$g({\bf r})$,

\begin{eqnarray}
\rho_{ch}(\mathbf{r}) = \int e\rho_p(\mathbf{r'})g(\mathbf{r}-\mathbf{r'})d\mathbf{r'}\\
g(\mathbf{r}) = (a\sqrt{\pi})^{-3}\exp(-r^2/a^2) \end{eqnarray}
with $a=0.8$ fm.  

Experimental charge radii values are from Angeli\cite{radii}.
The results show that RMF can describe the 
charge radii of these nuclei with sufficient accuracy. One sees that in most 
of the nuclei, the difference between measurement and theory is less than 1\%.

\begin{figure}
\resizebox{!}{!}{\includegraphics{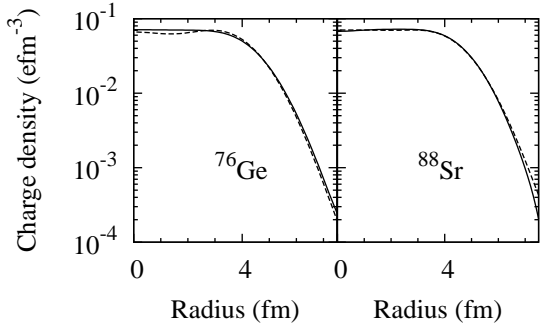}}
\caption{Comparison of charge density obtained from Fourier-Bessel analysis of 
experimental electron scattering data (solid line) and calculated in the present work (dashed line).
\label{chden}}
\end{figure}

Direct comparison of charge density is more difficult in absence of accurate
experimental information. De Vries \etal\cite{chden} have presented the 
coefficients of Fourier-Bessel expansion for charge density of a number on 
nuclei 
extracted from electron scattering 
data.  It includes two nuclei of our interest, $^{76}$Ge and $^{88}$Sr. 
In Fig. \ref{chden}, we compare the charge density extracted from the 
Fourier-Bessel coefficients and our calculated results for the above two 
nuclei. One can see that the theoretical and experimental values agree very
well, particularly at larger radii values, which is the region expected to
contribute to the optical potential at low projectile energy. 
However, in absence of information on error in the density values, this
conclusion can remain only tentative. 

Next, we compare the results for the reaction calculation in the above 
mentioned reactions with experiments. As the astrophysically 
important Gamow
window lies in the region 1.3 to 3.9 MeV for these nuclei, we present the 
results covering this energy region. 
The cross-section  varies very rapidly at such low energy making comparison 
between theory and experiment rather difficult. The usual
practice in low energy nuclear reaction is to compare another key observable,
{\em viz.} S-factor. The expression of the astrophysical S-factor\cite{6080} is given by, 
\begin{equation}
S(E)=E\sigma(E)e^{2\pi\eta}
\end{equation}
where E is the energy in centre of mass frame in KeV, $\sigma(E)$ is reaction
cross-section in barn and $\eta$ indicates
 the Sommerfeld parameter which may be obtained from the relation,
$2\pi\eta=31.29 Z_{p}Z_{t}\sqrt{\mu/E}$. 
Here, $Z_{p}$ and $Z_{t}$ are the charge numbers of the projectile and the
target, respectively and 
$\mu$ is the reduced mass (in amu). It varies much slowly than reaction
cross-sections as the exponential energy dependence of cross-section
is not present in it. For this reason, we calculate this quantity
and compare it with experimentally extracted values.

Figures 2 and 3 show the results for the reactions $^{84,86-88}$Sr($p,\ga$). 
The results compare favourably with experiments compared to the NON-SMOKER code
calculations of Rauscher \etal\cite{Smoker}.
However, it needs to be pointed out that, in the case of $^{87}$Sr, 
theoretical results overpredict the cross section values. It 
was suggested\cite{Sr1} that perhaps the agreement with theory (in their
case the NON-SMOKER calculation) worsens as one goes to more neutron-rich
nuclei. However, as one can see in the right panel of Figure 2, this trend
is not shared by the present calculation.


Figure 3 shows the results for the ($p,n$) reactions on (a) $^{75}$As, 
(b) $^{76}$Ge and (c) $^{85}$Rb targets.  These reactions
(along with their inverse reactions) are listed among the ten most important 
reactions in deciding the abundance of the $p$ nuclei in Rapp \etal\cite{astro1}. 
The three measurements for the $^{75}$As($p,n$) reaction are rather old and 
error values are not available for most of the measurements. The quoted error 
in cross section is 10\% or above. In $^{76}$Ge, we find that the calculation
systematically overpredict the results by as much as 60\%.  
On the other hand, the calculations for the  $^{85}$Rb($p,n$) reaction
produce excellent match with experimental measurements.

We find that our calculation can reproduce the S-factor values with reasonable
success. Even in the worst case, calculation is off by a factor less than two
while the cross section values range over four orders of magnitude. 
However, one should remember that in astrophysical calculations, the rates are
often varied by a large factor, {\em viz.} ten or hundred\cite{astro1}. Thus 
the present microscopic calculations can be used to obtain rates which are
dependable for astrophysical calculations.

We point out that in our earlier work\cite{NiCu}, we 
have seen that the default local  and global optical potentials
\cite{omp} in the TALYS package also can be used with suitable normalization of
gamma ray strength to produce comparable results for certain energy ranges. 
In the present case also, suitable selection of the parameter brings the values calculated with default potential close to experimental values.
However, we believe that the present microscopic approach is more suitable as 
no normalization is necessary and the method can be extended to reactions where experimental data are not available.

In summary, cross sections for low energy ($p,\ga$)  and ($p,n$) reactions for 
a number of nuclei in mass 80 region in the energy regime important for 
explosive nucleosynthesis have been calculated using the TALYS code. The microscopic optical potential has been 
obtained by folding the DDM3Y microscopic interaction with the nuclear 
densities obtained from RMF calculation using the Lagrangian density
FSU Gold. 

This work has been carried out with financial assistance of the UGC sponsored
DRS Programme of the Department of Physics of the University of Calcutta.
CL acknowledges the grant of a fellowship awarded by the UGC.
GG gratefully acknowledges the hospitality of the ICTP, 
Trieste where a part of the work was carried out.

\newpage

\begin{table}
\caption{Experimental binding energy and charge radii values 
compared with calculated results.} 
\begin{center}
\begin{tabular}{lcccc}\hline
&\multicolumn{2}{c}{B.E.(MeV)}
&\multicolumn{2}{c}{$r_{ch}$(fm)} \\
& Exp.&Theo.& Exp.&Theo.\\\hline
$^{84}$Sr &728.90&727.53& 4.236&4.232  \\
$^{86}$Sr&748.93 &748.27&4.226&4.240   \\
$^{87}$Sr& 757.36&757.17&4.220&4.245  \\
$^{88}$Sr& 768.47&768.47*&4.220&4.249  \\
$^{75}$As&652.56 &652.38&4.097&4.082 \\ 
$^{76}$Ge&661.60 &660.69&4.081&4.053\\
$^{85}$Rb&739.28&738.70&4.203&4.218 \\ 
\hline
\end{tabular}
\end{center}
$^*$ Normalized following the prescription of \cite{rmfcor,rmfcor1}.
\end{table}

\newpage

\begin{figure}
\resizebox{!}{!}{\includegraphics{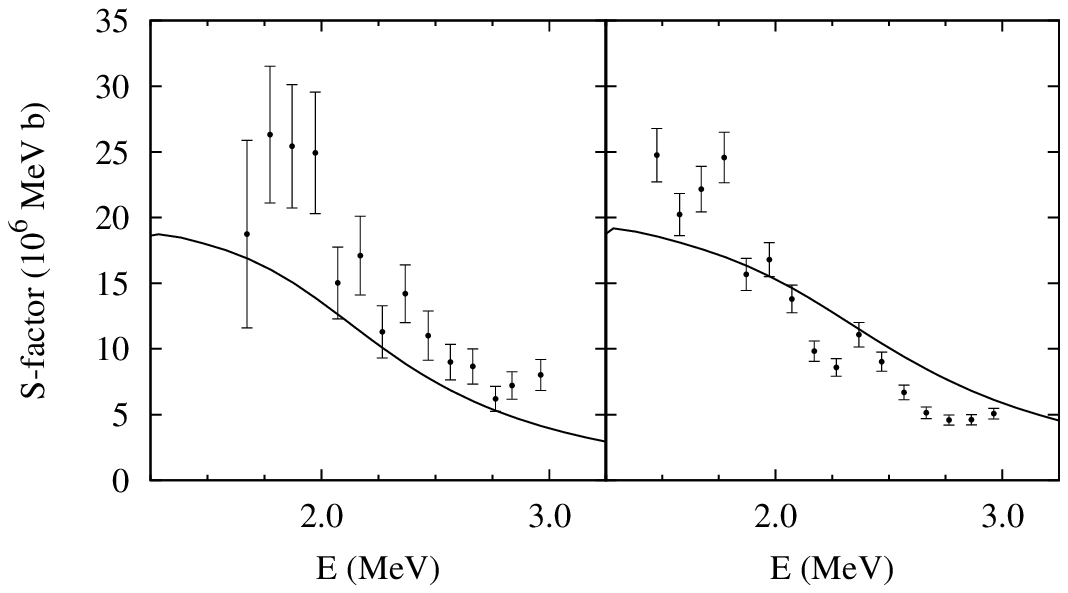}}
\caption{Experimental and calculated S-factors for ($p,\ga$) reactions in (a)$^{84}$Sr and
(b) $^{86}$Sr targets.}  
\end{figure}
\begin{figure}
\resizebox{!}{!}{\includegraphics{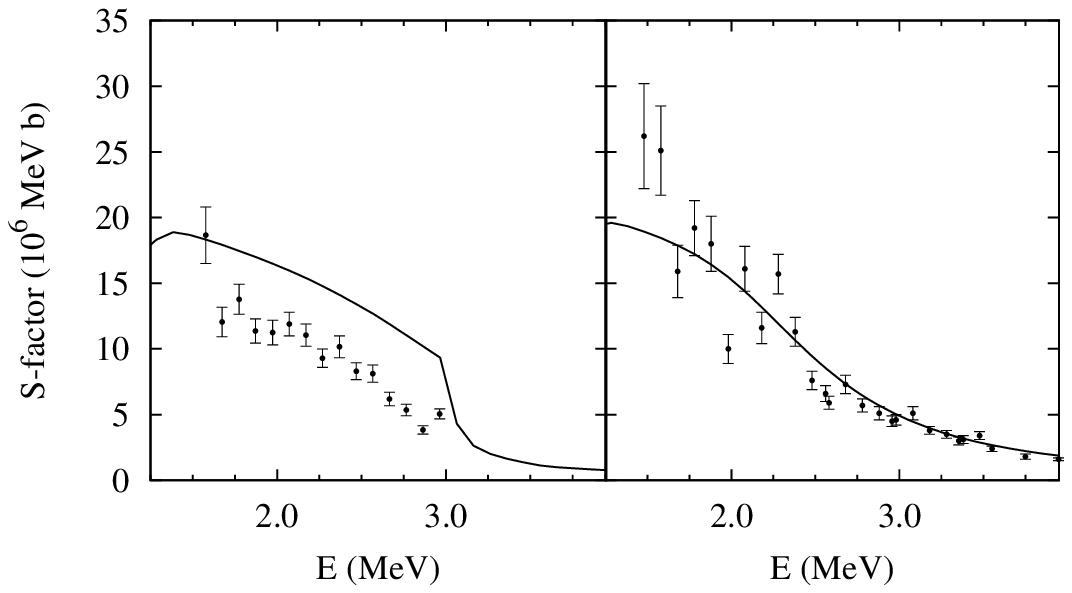}}
\caption{Experimental and calculated S-factors for ($p,\ga$) reactions in (a)$^{87}$Sr and
(b) $^{88}$Sr targets.}
\end{figure}
\begin{figure}
\resizebox{10cm}{!}{\includegraphics{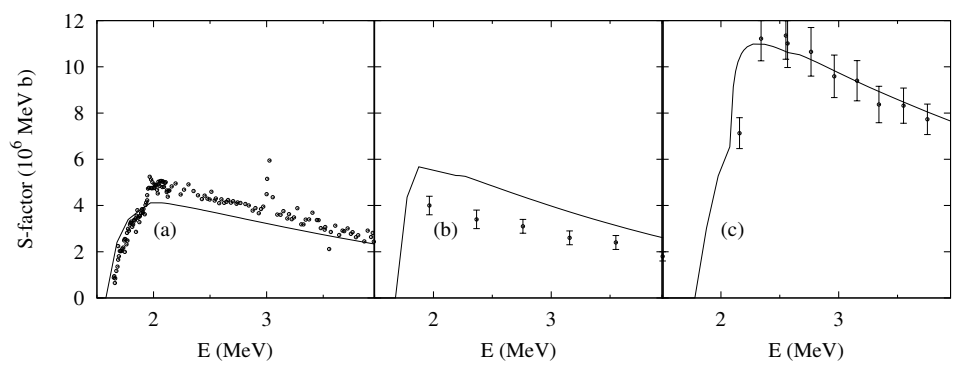}}
\caption{Experimental and calculated S-factors for (a) $^{75}$As($p,n$), (b) $^{76}$Ge($p,n$) and 
(c) $^{85}$Rb($p,n$) 
reactions, respectively.}
\end{figure}

\end{document}